\documentclass[twocolumn]{aastex61}

\shorttitle{ALMA observations of a young lensed starburst at $z=1.7$}
\shortauthors{Gonz\'alez-L\'opez, Barrientos et al.}

\begin{document}

\title{ALMA resolves the molecular gas in a young low-metallicity starburst galaxy at $z=1.7$}
\correspondingauthor{Jorge Gonz\'alez-L\'opez}
\email{jgonzal@astro.puc.cl}
\author[0000-0003-3926-1411]{Jorge Gonz\'alez-L\'opez}
\affil{Instituto de Astrof\'{\i}sica and Centro de Astroingenier{\'{\i}}a, Facultad de F\'{i}sica, Pontificia Universidad Cat\'{o}lica de Chile, Casilla 306, Santiago 22, Chile}
\author{L. Felipe Barrientos}
\affil{Instituto de Astrof\'{\i}sica and Centro de Astroingenier{\'{\i}}a, Facultad de F\'{i}sica, Pontificia Universidad Cat\'{o}lica de Chile, Casilla 306, Santiago 22, Chile}
\author{M.~D. Gladders}
\affil{Dept. of Astronomy and Astrophysics, Univ. of Chicago, 5640 S. Ellis Ave., Chicago, IL, 60637\\}
\affil{Kavli Institute for Cosmological Physics at The University of Chicago, Chicago, IL, USA}
\author{Eva Wuyts}
\affil{Max-Planck-Institut f{\"u}r extraterrestrische Physik, Giessenbachstr. 1, D-85741 Garching, Germany}
\author{Jane Rigby}
\affil{NASA Goddard Space Flight Center, Greenbelt, MD, USA}
\author{Keren Sharon}
\affil{Department of Astronomy, University of Michigan, 1085 S. University Avenue, Ann Arbor, MI 48109, USA}
\author{Manuel Aravena}
\affil{N\'ucleo de Astronom\'{\i}a, Facultad de Ingenier\'{\i}a y Ciencias, Universidad Diego Portales, Av. Ej\'ercito 441, Santiago, Chile}
\author{Matthew B. Bayliss}
\affil{MIT Kavli Institute for Astrophysics and Space Research, 77 Massachusetts Avenue, Cambridge, MA 02139, USA}
\author{Eduardo Ibar}
\affil{Instituto de F\'{\i}sica y Astronom\'{\i}a, Universidad de Valpara\'{\i}so, Avda. Gran Breta\~na 1111, Valpara\'{\i}so, Chile}

\begin{abstract}
We present Atacama Large Millimeter/submillimeter Array observations of CO lines and dust continuum emission of the source RCSGA 032727--132609, a young $z=1.7$ low-metallicity starburst galaxy. The CO(3-2) and CO(6-5) lines, and continuum at rest-frame $450\,\mu m$ are detected and show a resolved structure in the image plane. 
We use the corresponding lensing model to obtain a source plane reconstruction of the detected emissions revealing intrinsic flux density of $S_{450\,\mu m}=23.5_{-8.1}^{+26.8}$ $\mu$Jy and intrinsic CO luminosities $L'_{\rm CO(3-2)}=2.90_{-0.23}^{+0.21}\times10^{8}$ ${\rm K\,km\,s^{-1}\,pc^{2}}$ and $L'_{\rm CO(6-5)}=8.0_{-1.3}^{+1.4}\times10^{7}$ ${\rm K\,km\,s^{-1}\,pc^{2}}$. We used the resolved properties in the source plane to obtain molecular gas and star-formation rate surface densities of  $\Sigma_{\rm H2}=16.2_{-3.5}^{+5.8}\,{\rm M}_{\odot}\,{\rm pc}^{-2}$ and $\Sigma_{\rm SFR}=0.54_{-0.27}^{+0.89}\,{\rm M}_{\odot}\,{\rm yr}^{-1}\,{\rm kpc}^{-2}$ respectively. The intrinsic properties of RCSGA 032727--132609 show an enhanced star-formation activity compared to local spiral galaxies with similar molecular gas densities, supporting the ongoing merger-starburst phase scenario. RCSGA 032727--132609 also appears to be a low--density starburst galaxy similar to local blue compact dwarf galaxies, which have been suggested as local analogs to high-redshift low-metallicity starburst systems.
Finally, the CO excitation level in the galaxy is consistent with having the peak at ${\rm J}\sim5$, with a higher excitation concentrated in the star-forming clumps. 
\end{abstract}


\section{Introduction}\label{sec:intro}

The study of the distribution of molecular hydrogen (H2) and star formation rate (SFR) in high redshift galaxies has been a prolific field for the last 20 years. The CO emission lines, good tracers of the molecular gas \citep{Omont2007}, and the cold dust continuum emission have been resolved in a number of galaxies at $z>1$ \citep{Carilli_Walter2013,Hodge2015}. 
These observations have shown a picture where the SFR is strongly linked to the distribution of molecular gas, indicating that two mechanisms for star formation could be in place, the quiescent, normal phase and the starburst phase \citep{Daddi2010,Genzel2010,Genzel2015}. Most galaxies at all redshifts appear to be in the main sequence (MS) phase \citep[e.g.][]{Brinchmann2004}, where the SFR is proportional to the stellar and molecular gas masses. These galaxies would be producing stars following the standard mechanism (still not fully understood) of star formation. The starburst phase corresponds to those galaxies that, for a given stellar and/or molecular gas mass, have SFRs of at least 10 times higher than that of MS galaxies. The starburst galaxies show a more efficient process of star formation, expected to be triggered by other factors such as galaxy interactions or particular environmental conditions \citep{Dekel2009,Engel2010}. 

It has been found that starburst galaxies at high redshift ($z\geq1$) show similar star formation conditions and efficiencies as those observed in particular regions of local (ultra) luminous infrared galaxies \citep{Carilli_Walter2013}. 
The unresolved nature of most of the high redshift observations does not allow us to see whether the high redshift starburst galaxies have higher efficiencies across the whole galaxy or the star formation is dominated by small, highly efficient regions as in the local galaxies. The need for higher resolution observations is clear if we want to understand the starburst process and its conditions. 

The best way to resolve the emission in galaxies without increasing substantially the observation time is by targeting gravitationally lensed galaxies, specially the spatially resolved bright gravitational arcs. Observations of several lensed sources already show that different star forming regions show different star formation conditions \citep{Hezaveh2013,ALMA2015,Hatsukade2015,Spilker2016}. In this letter we present the CO and continuum emission of the second-brightest optical giant arc discovered to date, the arc RCSGA 032727--132609 (hereafter RCS0327), strongly lensed by the foreground galaxy cluster RCS2 032727--132623 at z = 0.564 \citep{Wuyts2010}, to reveal the star formation process in a starburst at $z=1.7$. The kinematic analysis of H$\alpha$ on RCS0327 strongly suggests an interaction consistent with a merger of galaxies \citep{Wuyts2014}. The SFRs measured in individual clumps and across the galaxy fall well above the MS relation for galaxies at $z=1.7$ and are consistent with a young low-metallicity ($\approx0.3{\rm Z}_{\odot}$) starburst enhanced by the ongoing merger \citep{Wuyts2014,Whitaker2014}. Furthermore, \citet{Bordoloi2016} measured resolved galactic winds in RCS0327, showing that the outflows are comparable to those observed in local starbursts. 

Throughout this paper, we adopt a cosmology with $H_{0} = 67.8$ km s$^{-1}$ Mpc$^{-1}$, $\Omega_{\rm m}=0.308$ and $\Omega_{\rm \Lambda}=0.692$ \citep{Planck2016}.

\section{Observations}\label{sec:obs}

The target RCS0327 was observed as part of the ALMA project 2015.1.00920.S which aimed to detect the emission lines CO(3-2), CO(6-5) and CO(8-7) and the underlying continuum. At the redshift for RCS0327 of $z = 1.7037455 \pm 0.000005$ \citep{Wuyts2014} the lines are centered at 127.895 GHz for CO(3-2), 255.746 GHz for CO(6-5) and at 340.934 GHz in for CO(8-7). The observations of the lines CO(6-5) and CO(3-2) were carried out during January 1st and 10th 2016. The observations of CO(8-7) unfortunately were not completed during cycle 3. 

In each observation, two overlapping spectral windows (SPWs) were placed to detect the line and two were placed to detect continuum. The passband and amplitude calibrator for both observations was J0423-0120, with J0336-1302 being used as phase calibrator. The reduction of the data was performed using the scripts provided by ALMA and used the pipeline version for cycle 3 and {\sc casa} \citep{Mcmullin2007} version 4.5.1. 

Both emission lines were observed using the array configuration C36-1. The imaging of the calibrated data was done using the {\sc casa} task \texttt{clean}. The data for CO(3-2) was imaged using natural weighting returning a synthesized beam size of 2\farcs54$\times$1\farcs85 and position angle of -85\fdg68. The natural weighting synthesized beam for the CO(6-5) observations was 1\farcs49$\times$0\farcs98 and position angle of 79\fdg75. To achieve a beam size similar to CO(3-2), a taper is applied to the CO(6-5) data to obtain a synthesized beam of 2\farcs16$\times$2\farcs08 at a position angle of 59\fdg64. 

Data cubes using a spectral resolution of 20 km s$^{-1}$ were created for each of the emission lines. The continuum emission is estimated by using the line free channels and then is subtracted from the $uv$ data using \texttt{uvcontsub} and new continuum subtracted cubes are created using the same procedure as before. The continuum and line images were interactively cleaned by manually masking out the emission. 

\section{Results}\label{sec:results}

\begin{figure*}
\gridline{\fig{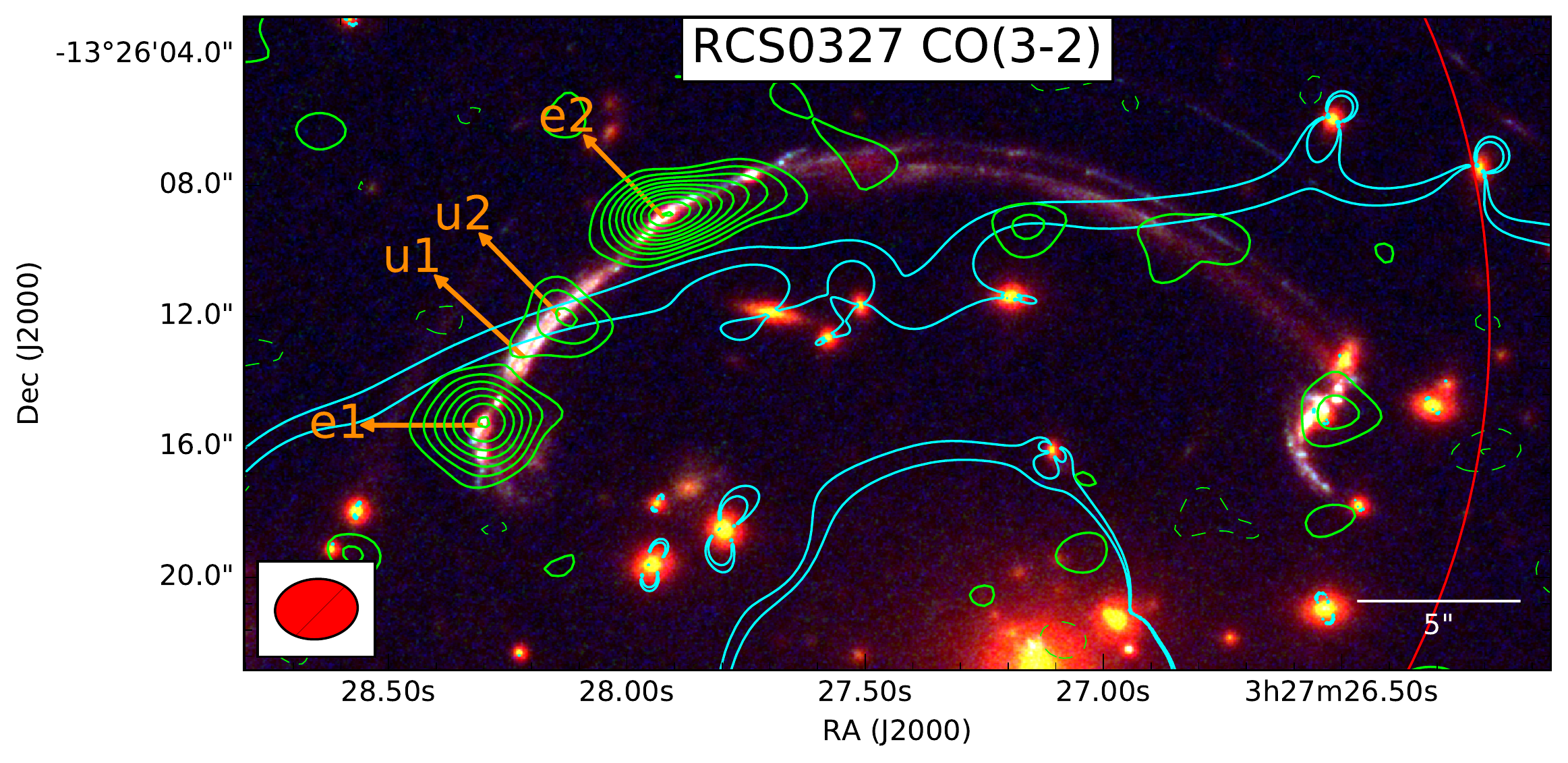}{0.5\textwidth}{(a)}
          \fig{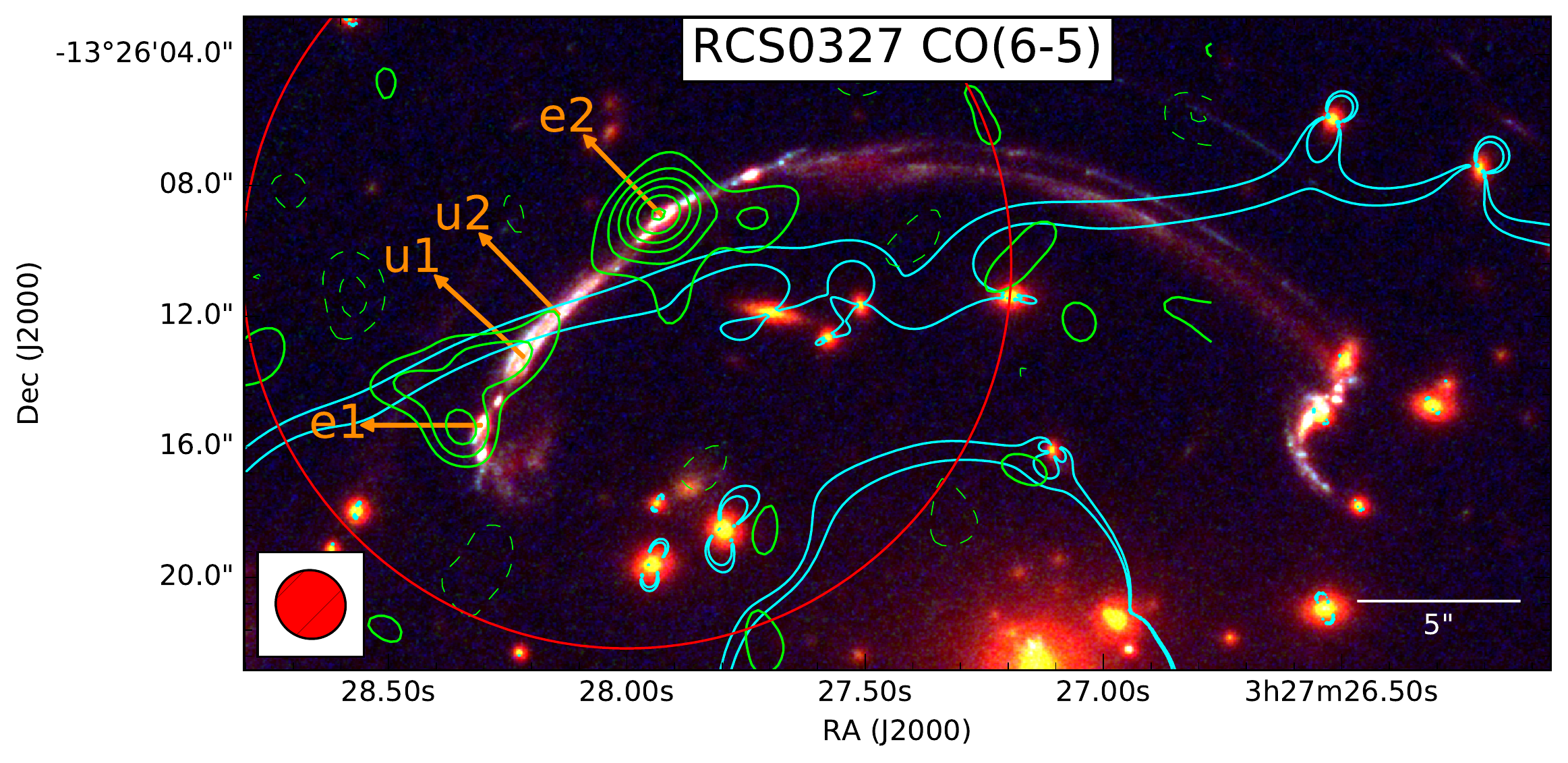}{0.5\textwidth}{(b)}}
\gridline{\fig{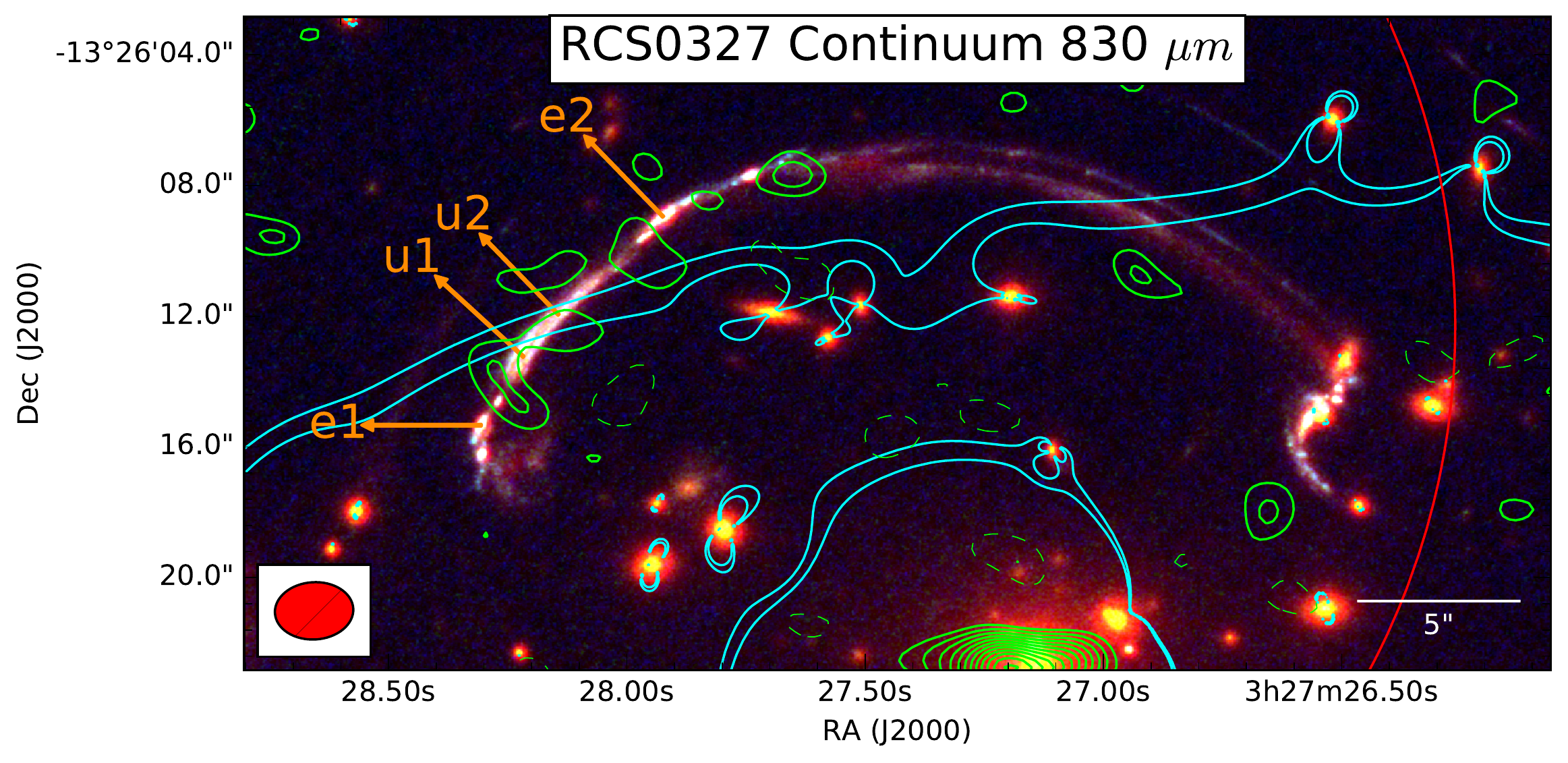}{0.5\textwidth}{(c)}
          \fig{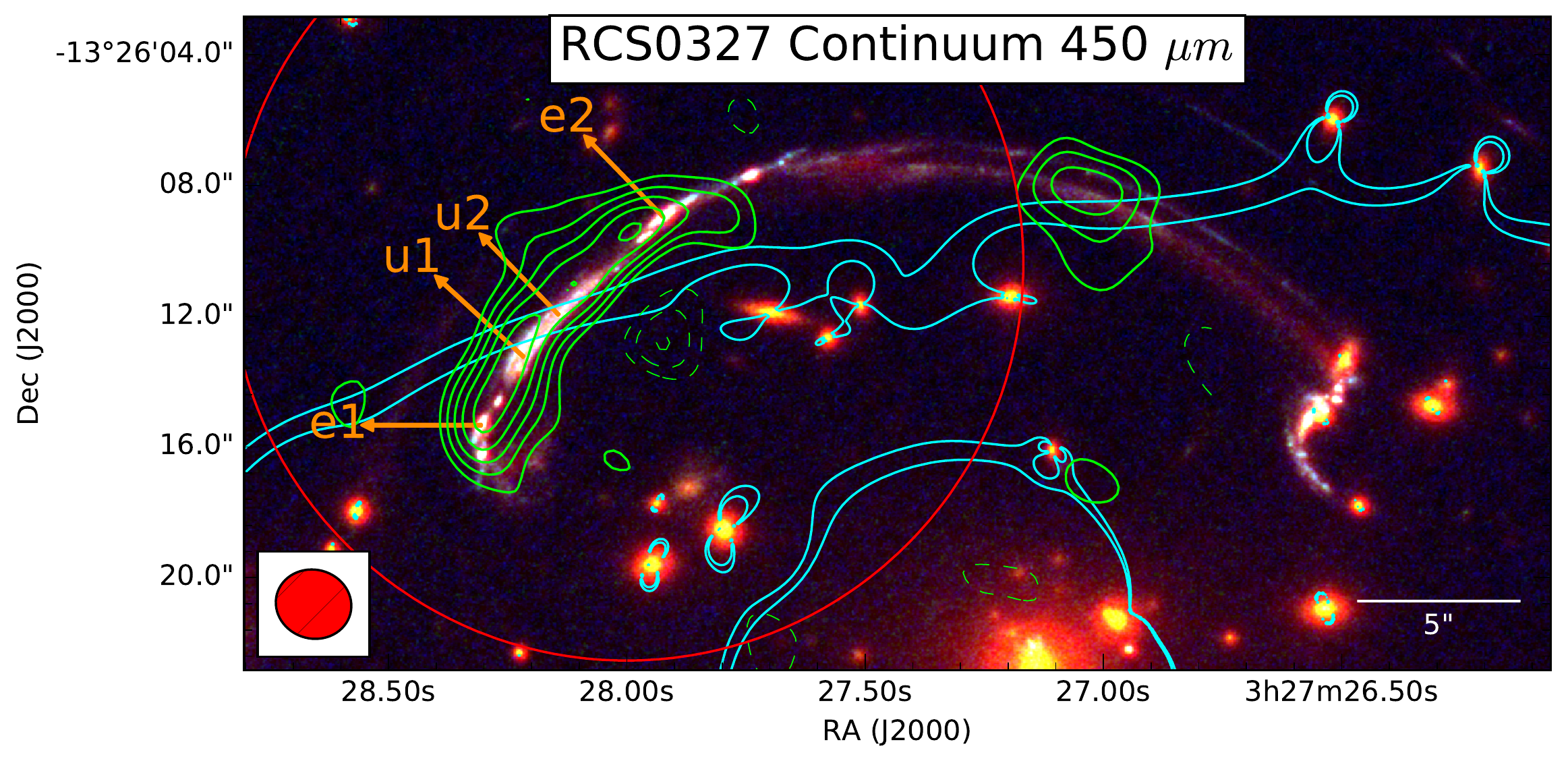}{0.5\textwidth}{(d)}}
\gridline{\fig{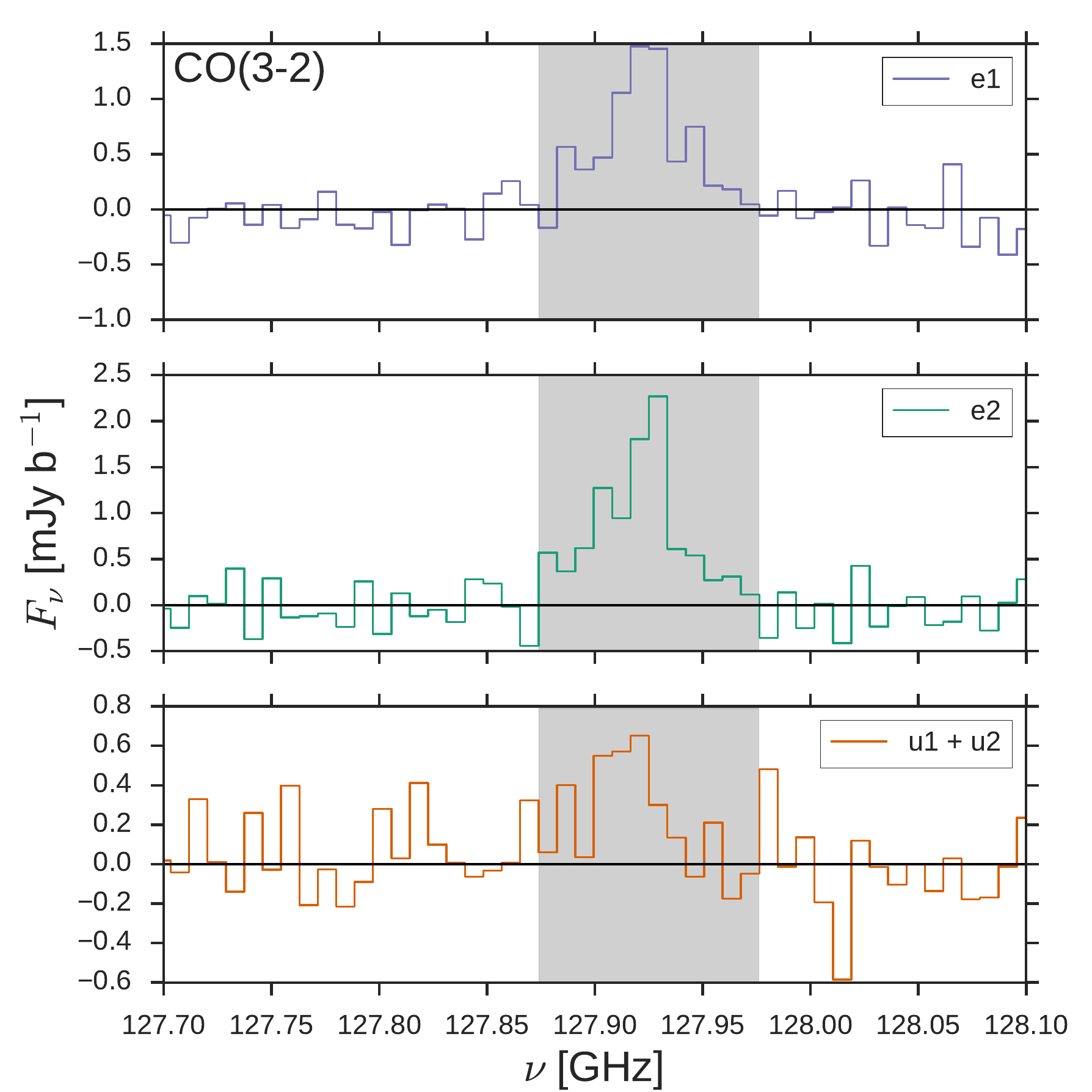}{0.5\textwidth}{(e)}
          \fig{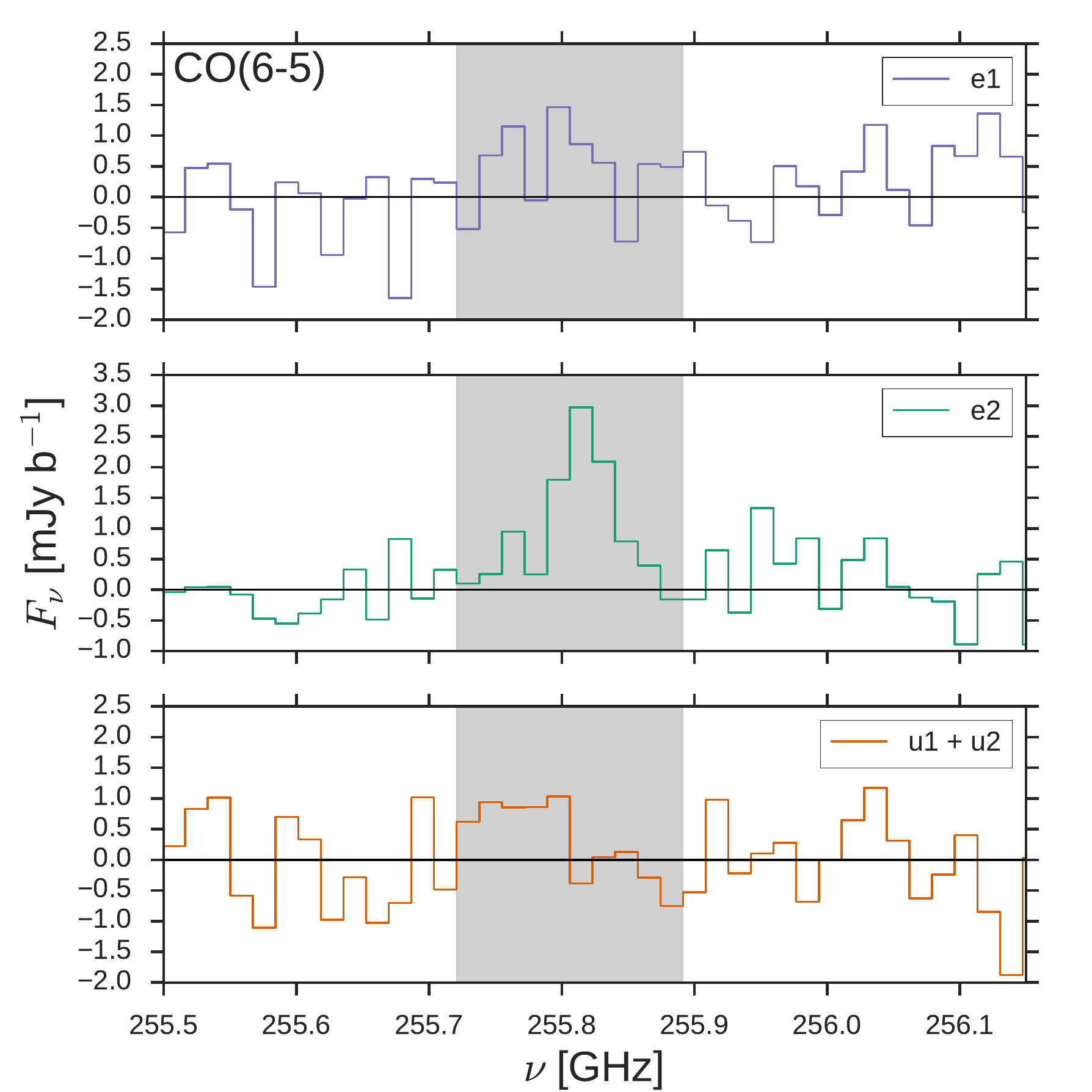}{0.5\textwidth}{(f)}}
\caption{{\it panels a, b, c and d:} 2D maps representation of the lines and continuum emission over a color image created using {\it HST} imaging (F390W for blue, F606W and F814W for green and F098M, F125W and F160W for red). The contours represent $1\sigma$ steps of the emissions starting at $\pm2\sigma$. The red lines shows the primary beam for both observations while the synthesized beams are shown in the left corner of each panel. The labeled regions correspond to those identified in \citet{Sharon2012}. The cyan solid lines represent the critical curves for a source at $z=1.704$.
{\it panels e and f:} CO(3-2) and CO(6-5) spectra extracted in different regions of the arc labeled in upper panels. The gray regions shows the frequency range used to create the line emission image.\label{fig:2D_maps_spectra} }
\end{figure*}

The CO(3-2), CO(6-5) and rest-frame $450\,\,\mu m$ continuum emission are detected in the brightest regions of the arc (Figure \ref{fig:2D_maps_spectra}). The continuum emission at rest-frame $830\,\,\mu m$ is also detected but with lower significance. 
To find the CO(3-2) total flux we extract the spectra on the positions e1, e2 and u1+u2, clumps identified by \citet{Sharon2012} that are near the peaks of the observed emission, which are plotted in the bottom panel of Figure \ref{fig:2D_maps_spectra}. A small astrometric offset was applied to the {\it HST} coordinates to match the $830\,\,\mu m$ continuum detection of the brightest cluster galaxy to its NIR counterpart. We find that the region going from 127.874--127.976 GHz ($\approx239.3$ km s$^{-1}$) provides a good range for the total emission of the line. The CO(3-2) line flux measured in the image plane for the section of the arc going from e1 to e2 (corresponding to lensed images 1 and 2) is ${I_{\rm CO(3-2)}}=0.557\pm 0.071$ Jy km s$^{-1}$ while for CO(6-5) is ${I_{\rm CO(6-5)}}=0.885\pm 0.117$ Jy km s$^{-1}$.

The spatially integrated continuum emission measured in the same region as CO(3-2) is ${S_{430\,\mu m}}=960 \pm 98 \,\mu {\rm Jy}$ and ${S_{875\,\mu m}}=141 \pm 31 \,\mu {\rm Jy}$. A second emitting region is detected over the arc, outside the primary beam (PB), marked with a red circle in  the middle panel of Figure \ref{fig:2D_maps_spectra}. The continuum emission for that component is $S_{430\,\mu m,\, 2nd}=737 \pm 370 \,\mu {\rm Jy}$. 

\section{Discussion}

\subsection{Source plane reconstruction}

\begin{figure*}
\epsscale{1.1}
\plotone{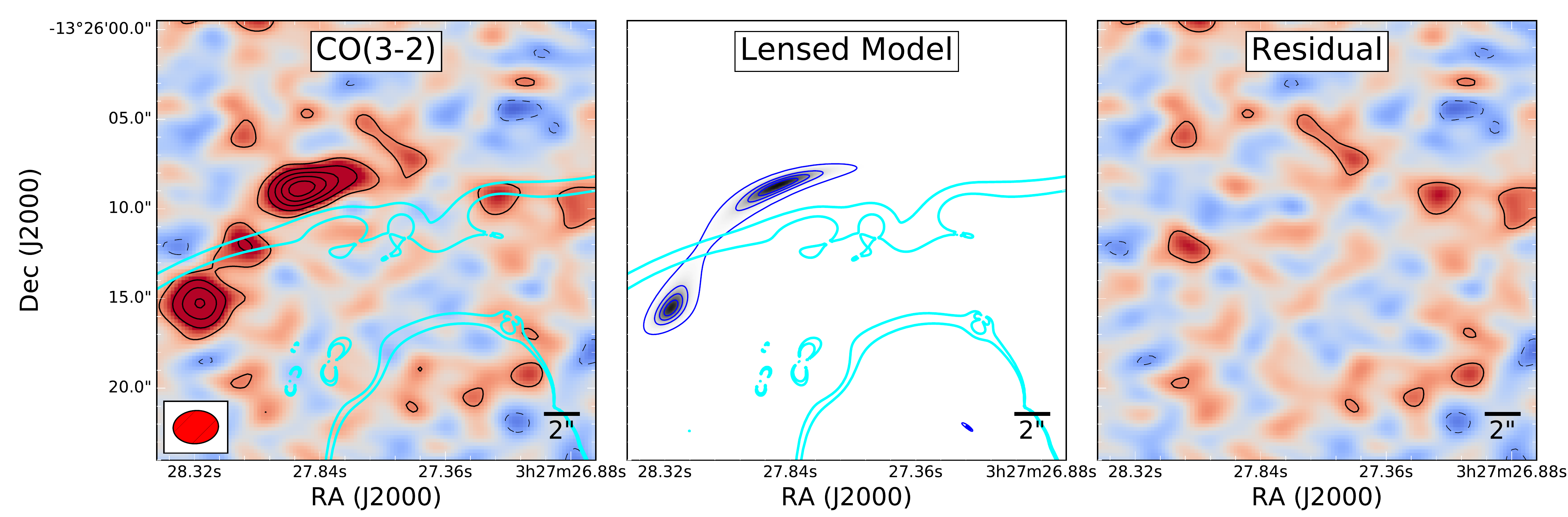}
\plotone{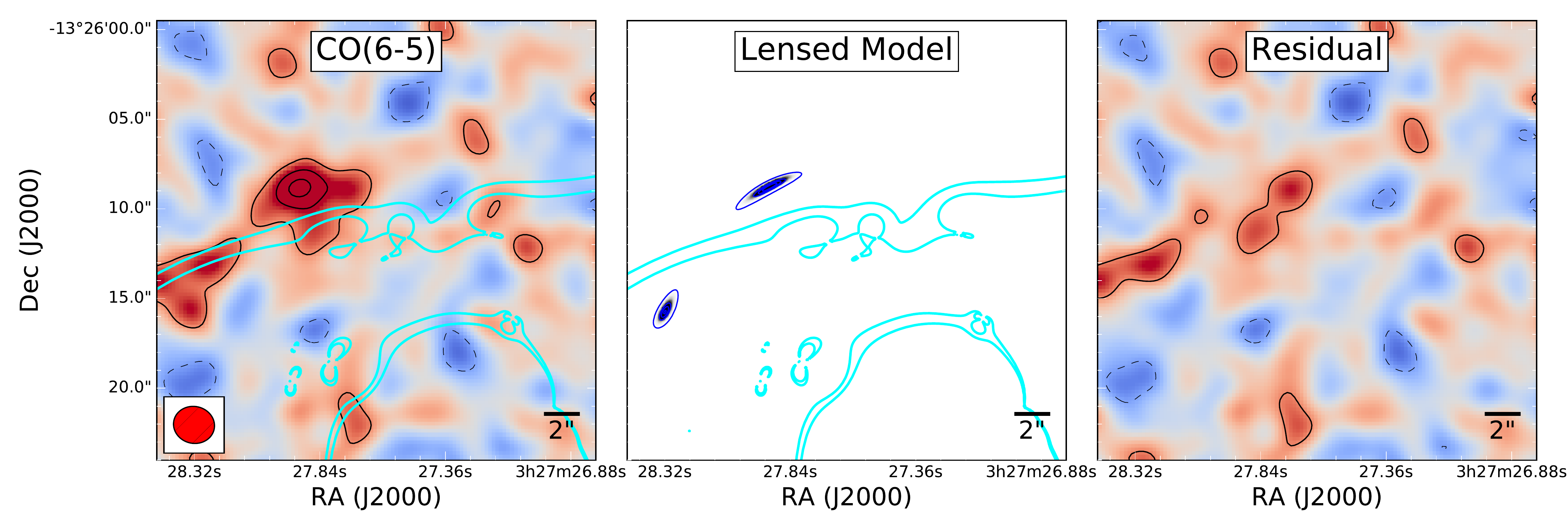}
\plotone{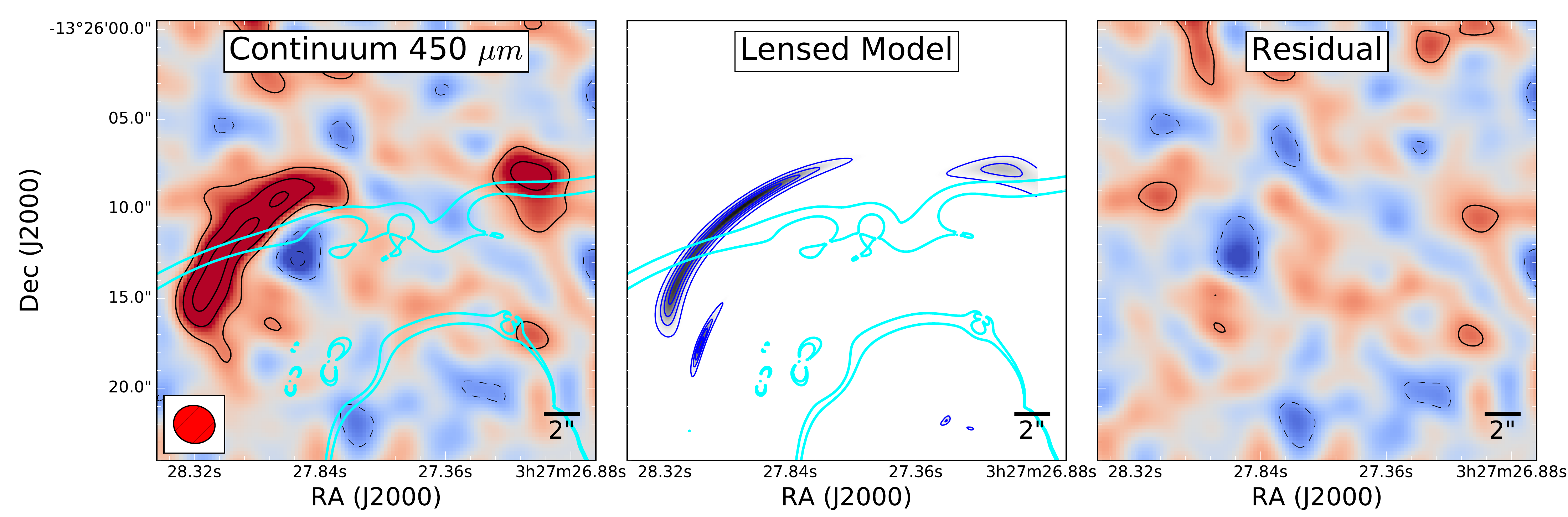}
\caption{Results from fitting the observed emission. In the left panels we show the observed emission for CO(3-2), CO(6-5) and continuum at $450\,\mu m$ with contour levels in steps of $2\sigma$ starting at $\pm2\sigma$. In the middle panels we show the image plane representation of the best fit model found in each case, the contours are 0.2, 0.4, 0.6, 0.8 and $1\times$ the peak of the model. The right panels show the residual images obtained after subtracting the best model simulated visibilities from the observed ones. The best model account for most of the emission in all cases leaving low significance residuals. The cyan solid lines represent the critical curves for a source at $z=1.7$. \label{fig:residual_maps}}
\end{figure*}

Working with a galaxy that is magnified by strong gravitational lensing requires additional analysis to recover the galaxy's intrinsic properties. Knowledge of the lensing deflection field is required to reconstruct the galaxy in the source-plane, and to account for the magnification--which can be highly spatially variable--. Different galaxy regions will be stretched and magnified by a different factor, resulting in different physical scales over the arc. To fully understand our detected emission we need to take them to the source plane, where the physical scale is unique. 

To do the source plane reconstruction for RCS0327, we used \texttt{uvmcmcfit},\footnote{\url{https://github.com/sbussmann/uvmcmcfit}} which is a \texttt{Python} implementation to fit emission models to interferometric data in the $uv$ plane \citep{Bussmann2013,Bussmann2015,Gonzalez-Lopez2017}. Exploring the emission in the $uv$ plane should return the maximum amount of information from the observations and it has proven to be useful in revealing the source plane emission in bright SMGs discovered by the South Pole Telescope and observed with ALMA \citep{Hezaveh2013,Spilker2016}. 

\texttt{Uvmcmcfit} can fit the source plane emission of a galaxy together with the lensing potential. The galaxy emission is fitted assuming a 2D elliptical Gaussian (special case of a S\'ersic profile with $n=0.5$) while the lensing potential is fitted by a singular isothermal ellipsoid (SIE). 
Recent high resolution imaging of high-redshift SMGs have shown no strong preference between fitting the dust continuum emission with Gaussian or S\'ersic profiles \citep{Hodge2016,Spilker2016}, supporting the usage of a simple 2d elliptical Gaussian function for the source model. 
We modified \texttt{uvmcmcfit} so that it uses a user provided lensing model deflection field allowing us to incorporate the detailed lensing model that is already available, based on HST imaging \citep{Sharon2012} and fit the ALMA data only to constrain the properties of the source plane emission. During the fitting of the emission, the lensing model is held fixed and only the source plane emission model is allowed to vary. The model found by using high-resolution {\it HST} observations of multiple images of strongly lensed galaxies at different redshifts together with the cluster members information outperforms in complexity and quality to the model we could find by fitting the lensing potential and source structure together using only the ALMA observations \citep{Sharon2012}.

\begin{figure*}[!t]
\epsscale{1.1}
\gridline{\fig{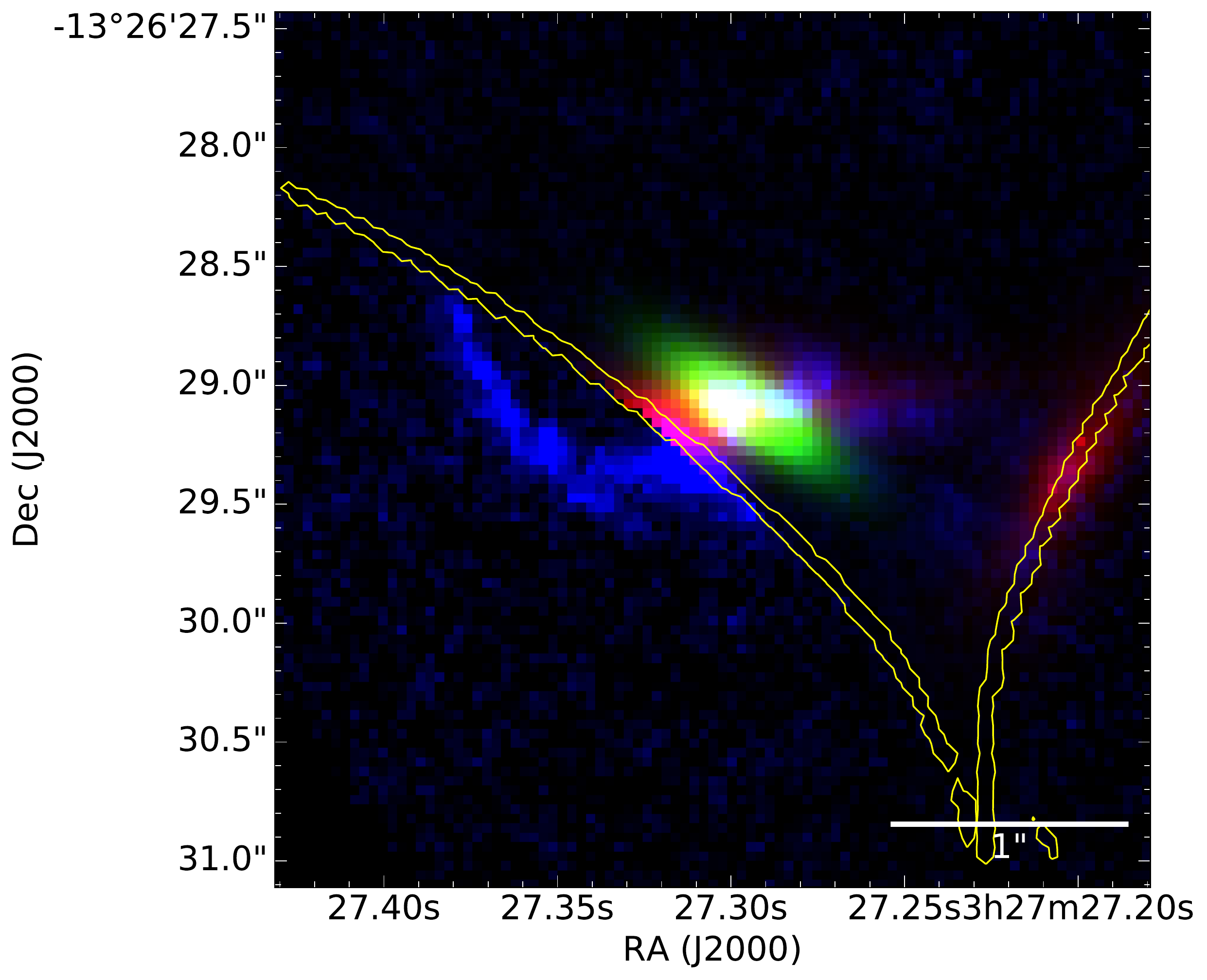}{0.5\textwidth}{(a)}
          \fig{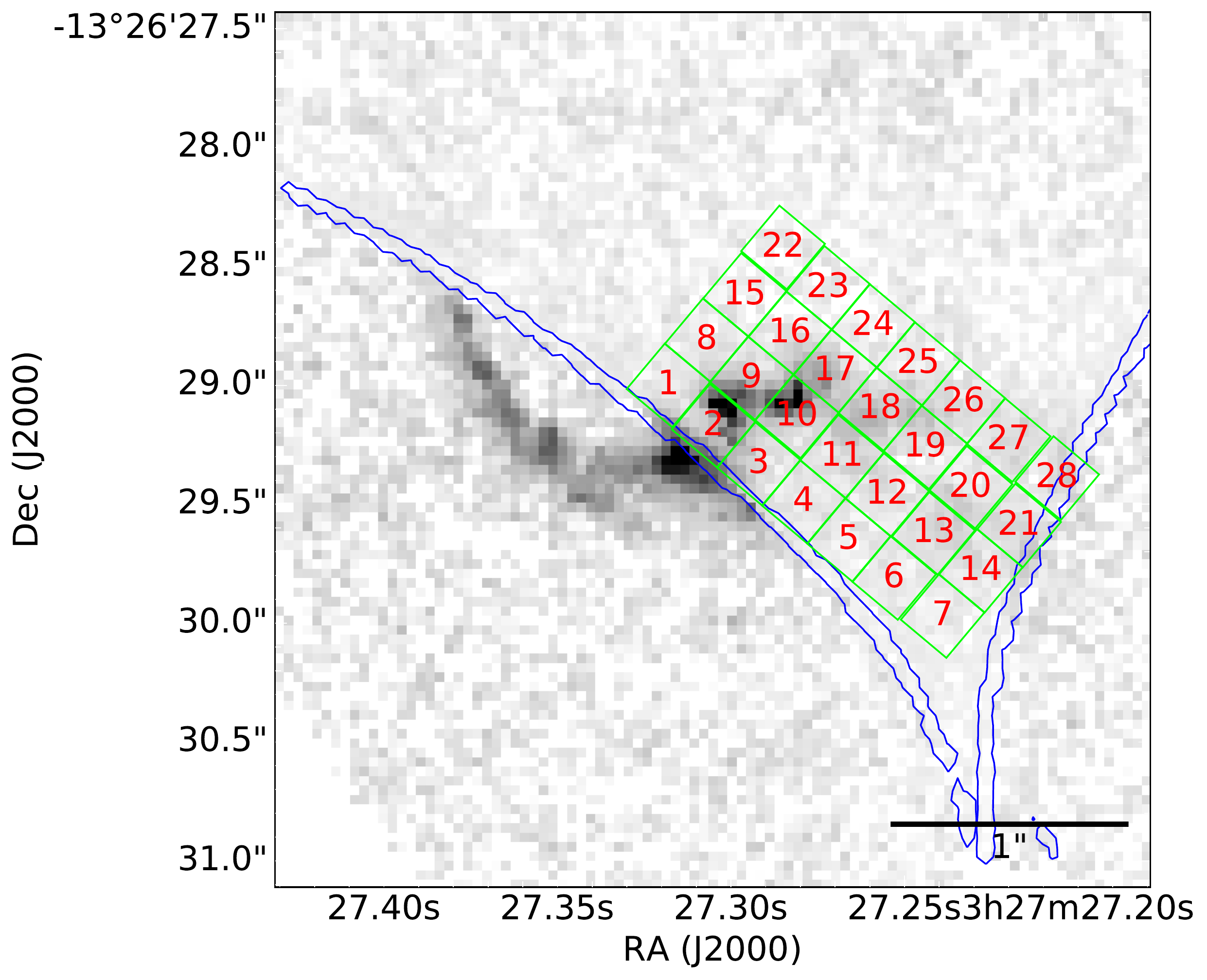}{0.5\textwidth}{(b)}}
\gridline{\fig{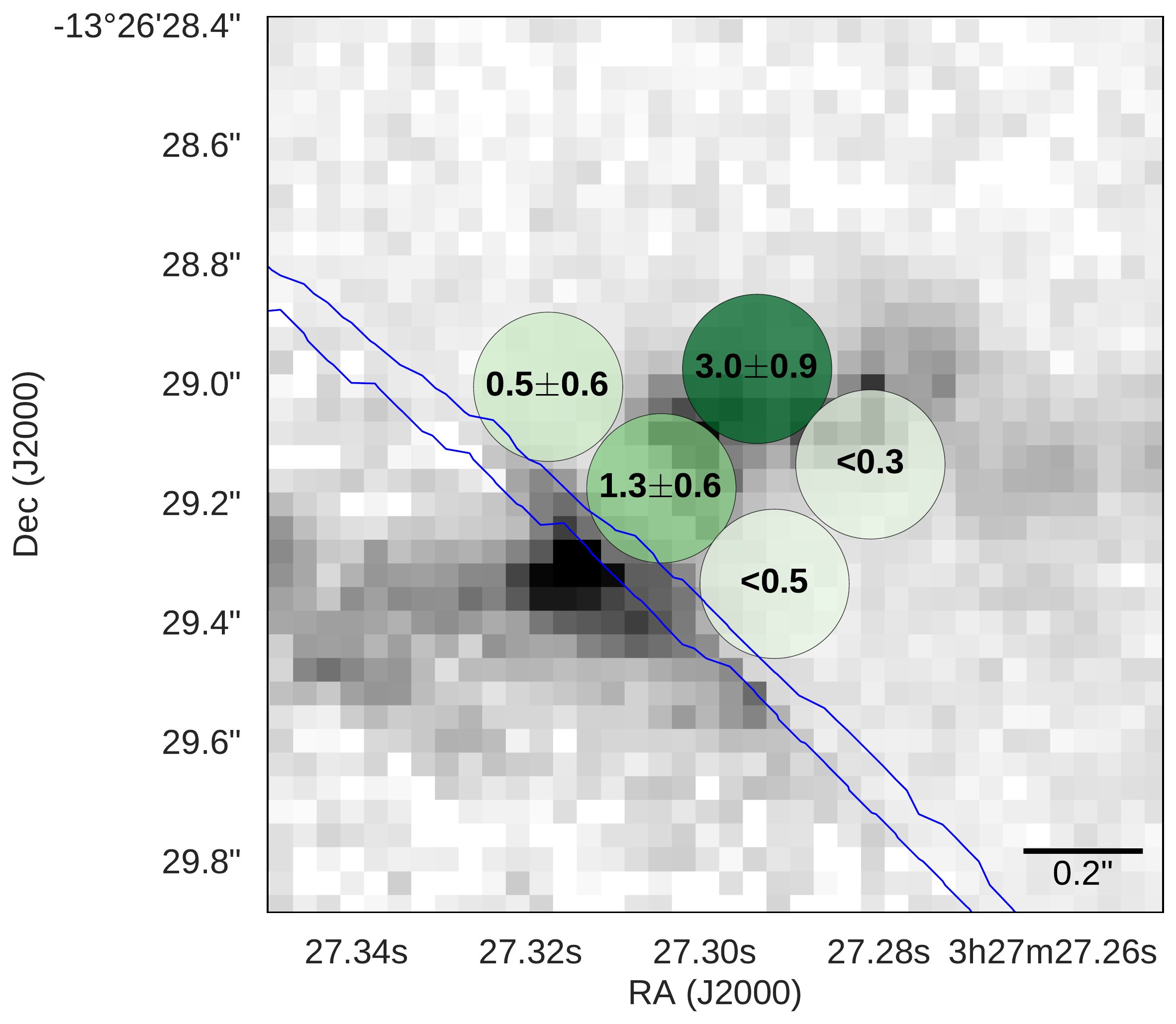}{0.5\textwidth}{(c)}}
\caption{The panel (a) shows a composite of the different observed emission of RCS0327 in the source plane. The blue channel represent the F390W emission (restframe UV), the green channel corresponds to the CO(3-2) emission and the red channels corresponds to the continuum emission at $450\,mu m$. The CO(3-2) and continuum emissions come mainly from the clumps labeled as a--f in \citet{Sharon2012} and \citet{Wuyts2014}. The panel (b) shows the different regions used to measure local gas surface density and SFR surface density, each region has a size of 0\farcs25 and cover the area of the galaxy that is better explorer by both CO(3-2) and continuum observations.  At $z=1.704$, 1\arcsec\, corresponds to $\approx8.684$ kpc. The cyan and blue solid lines represent the caustic curves in the source plane. The panel (c) shows a zoom in to the regions where CO(6-5) and CO(3-2) are detected and a constraint to the ratio CO(6-5)/CO(3-2) can be estimated. Darker circles correspond to larger values. \label{fig:source_plane}}
\end{figure*}

For the case of CO(3-2), a single 2D elliptical Gaussian was needed to fit the observed emission. We measure an intrinsic flux of $S_{\rm CO(3-2)}=70.5_{-5.6}^{+5.1}$ $\mu$Jy (for the frequency range of $\approx239.3$ km s$^{-1}$) and an effective radius $r_{\rm CO(3-2)}=0.142_{-0.027}^{+0.025}$ arcseconds ($1.23_{-0.23}^{+0.22}$ kpc).
The flux weighted magnification value for CO(3-2) is $\mu_{\rm CO(3-2)}=33.1_{-4.9}^{+5.0}$ and an intrinsic CO(3-2) luminosity of $L'_{\rm CO(3-2)}=2.90_{-0.23}^{+0.21}\times10^{8}$ ${\rm K\,km\,s^{-1}\,pc^{2}}$. 

In the case of CO(6-5), a single component gives $S_{\rm CO(6-5)}=77.7_{-12.9}^{+13.8}$ $\mu$Jy (for the frequency range of $\approx239.3$ km s$^{-1}$) and an effective radius $r_{\rm CO(6-5)}=0.055_{-0.018}^{+0.021}$ arcseconds ($0.48_{-0.16}^{+0.18}$ kpc) The flux weighted magnification value for CO(6-5) is $\mu_{\rm CO(6-5)}=47.4_{-9.2}^{+11.9}$ and an intrinsic CO(6-5) luminosity of $L'_{\rm CO(6-5)}=8.0_{-1.3}^{+1.4}\times10^{7}$ ${\rm K\,km\,s^{-1}\,pc^{2}}$.

To fit the continuum emission at $450\,\mu m$, we needed two Gaussians in the source plane, since a single component was not enough to account for the whole observed flux outside the PB in Figure \ref{fig:2D_maps_spectra}. The main component is well described by a Gaussian with $S_{450\,\mu m}=23.5_{-8.1}^{+26.8}$ $\mu$Jy and $r_{450\,\mu m}=0.147_{-0.035}^{+0.051}$ arcseconds ($1.28_{-0.30}^{+0.44}$ kpc). The second component returns $S_{b,450\,\mu m, 2nd}=25.5_{-10.6}^{+12.5}$ $\mu$Jy and $r_{b,450\,\mu m, 2nd}=0.17_{-0.07}^{+0.12}$ arcseconds.
The flux weighted magnification value for the main component of continuum emission is $\mu_{450\,\mu m}=38.5_{-20}^{+23.5}$, while for the second component is $\mu_{b,450\,\mu m}=28.1_{-15.3}^{+27.0}$. 

In Figure \ref{fig:residual_maps} we present the observed emission for CO(3-2), CO(6-5) and continuum at $450\,\mu m$ (left panels), the image plane representation of the best fit model found for each case (middle panels) and the residual images obtained after subtracting the best model simulated visibilities from the observed ones (right panels). In all cases the best fit models appear to account for most of the observed emission. 

\subsection{Gas and dust distribution}

\begin{deluxetable*}{cCCCCCC}
\tablecaption{Flux density values measured in the source plane.\label{tab:table}}
\tablehead{
\colhead{Region} & 
\colhead{$S_{450\,\mu m}$} & 
\colhead{$S_{\rm CO(3-2)}$\tablenotemark{a}} & 
\colhead{$S_{\rm CO(6-5)}$\tablenotemark{a}} & 
\colhead{$S_{\rm CO(6-5)}/S_{\rm CO(3-2)}$} & 
\colhead{$\Sigma_{\rm H2}$\tablenotemark{b}} & 
\colhead{$\Sigma_{\rm SFR}$}\\
\colhead{} & 
\colhead{$\mu {\rm Jy}$} & 
\colhead{$\mu {\rm Jy}$} & 
\colhead{$\mu {\rm Jy}$} &
\colhead{} &
\colhead{${\rm M}_{\odot}\,{\rm pc}^{-2}$} &
\colhead{${\rm M}_{\odot}\,{\rm yr}^{-1}\,{\rm kpc}^{-2}$} 
}
\colnumbers
\startdata 
Main Component\tablenotemark{c} & 48.1_{-16.6}^{+54.9} & 70.5_{-5.6}^{+5.1} & 77.1_{-12.9}^{+13.8} & 1.1\pm0.2& 16.2_{-3.5}^{+5.8} & 0.27_{-0.13}^{+0.43}\\
\hline
01 & 4.0\pm0.9 & 6.7\pm3.3 & 3.6\pm3.4 & 0.5\pm0.6& 7.1\pm3.5 & 0.16_{-0.04}^{+0.15}\\
02 (e, f, u) & 6.9\pm1.0 & 12.3\pm1.8 & 16.3\pm6.4 & 1.3\pm0.6& 13.0\pm1.9 & 0.27_{-0.04}^{+0.17}\\
03 & 1.4\pm0.6 & 6.6\pm2.1 & 0.0\pm1.6 & <0.5& 7.0\pm2.2 & 0.06_{-0.02}^{+0.10}\\
04 & 0.0\pm0.3 & 1.2\pm1.1 & 0.0\pm1.6 & \nodata & <2.3 & <0.10\\
05 & 0.0\pm0.2 & 0.1\pm0.7 & 0.0\pm1.5 & \nodata & <1.5 & <0.07\\
06 (t) & 0.0\pm0.2 & 0.0\pm0.7 & 0.0\pm1.6 & \nodata & <1.5 & <0.07\\
07 (t) & 0.0\pm0.3 & 0.0\pm0.7 & 0.0\pm1.8 & \nodata & <1.5 & <0.10\\
08 & 0.5\pm0.3 & 5.0\pm3.1 & 10.2\pm9.7 & \nodata & <6.6 & <0.10\\
09 (d) & 3.3\pm1.0 & 12.8\pm1.9 & 38.7\pm9.9 & 3.0\pm0.9& 13.5\pm2.0 & 0.13_{-0.04}^{+0.17}\\
10 (b, c) & 3.0\pm1.9 & 12.7\pm2.7 & 0.0\pm1.6 & <0.3& 13.4\pm2.9 & <0.64\\
11 (r) & 0.4\pm0.3 & 4.3\pm2.5 & 0.0\pm1.6 & \nodata & <5.3 & <0.10\\
12 (r) & 0.0\pm0.3 & 0.5\pm0.7 & 0.0\pm1.7 & \nodata & <1.5 & <0.10\\
13 (t) & 0.0\pm0.3 & 0.0\pm0.7 & 0.0\pm1.8 & \nodata & <1.5 & <0.10\\
14 & 0.0\pm0.4 & 0.0\pm0.7 & 0.0\pm2.1 & \nodata & <1.5 & <0.13\\
15 & 0.0\pm0.3 & 0.0\pm0.8 & 0.0\pm1.9 & \nodata & <1.7 & <0.10\\
16 & 0.3\pm0.3 & 0.1\pm0.7 & 0.0\pm1.8 & \nodata & <1.5 & <0.10\\
17 (a) & 0.9\pm0.6 & 0.2\pm0.7 & 0.0\pm1.8 & \nodata & <1.5 & <0.20\\
18 (r) & 0.7\pm0.7 & 0.1\pm0.7 & 0.0\pm1.9 & \nodata & <1.5 & <0.24\\
19 (r) & 0.1\pm0.3 & 0.0\pm0.7 & 0.0\pm1.9 & \nodata & <1.5 & <0.10\\
20 & 0.0\pm0.3 & 0.0\pm0.7 & 0.0\pm2.1 & \nodata & <1.5 & <0.10\\
21 & 0.2\pm0.5 & 0.0\pm0.7 & 0.0\pm2.8 & \nodata & <1.5 & <0.17\\
22 & 0.0\pm0.3 & 0.0\pm0.8 & 0.0\pm2.1 & \nodata & <1.7 & <0.10\\
23 & 0.0\pm0.3 & 0.0\pm0.8 & 0.0\pm2.0 & \nodata & <1.7 & <0.10\\
24 & 0.0\pm0.3 & 0.0\pm0.7 & 0.0\pm2.0 & \nodata & <1.5 & <0.10\\
25 & 0.1\pm0.3 & 0.0\pm0.7 & 0.0\pm2.1 & \nodata & <1.5 & <0.10\\
26 (s) & 0.1\pm0.3 & 0.0\pm0.7 & 0.0\pm2.2 & \nodata & <1.5 & <0.10\\
27 & 0.1\pm0.4 & 0.0\pm0.7 & 0.0\pm2.5 & \nodata & <1.5 & <0.13\\
28 & 0.4\pm0.5 & 0.0\pm0.8 & 0.0\pm3.1 & \nodata & <1.7 & <0.17\\
\enddata
\tablenotetext{a}{For the frequency range of $\approx239.3$ km s$^{-1}$.}
\tablenotetext{b}{Using $\alpha_{\rm CO}=0.8$.}
\tablenotetext{c}{ and source plane emission knots \citep{Sharon2012}.}
\tablecomments{Upper limits correspond to $2\sigma$.}
\end{deluxetable*}

We now use the CO(3-2) to estimate the amount and distribution of molecular gas in the galaxy. We first need to estimate the amount of CO(1-0) luminosity based on the observed CO(3-2) luminosity and use the CO conversion factor ($\alpha_{\rm CO}$) to convert it to molecular gas mass \citep{Carilli_Walter2013,Bolatto2013}. The CO excitation level depends mainly on the density and temperature of the gas and it has been found that the excitation level differ for different source populations. Similar results have been found for the CO conversion factor, where a typical value of $\alpha_{\rm CO}\sim0.8$ M$_{\odot}$ $({\rm K\,km\,s^{-1}\,pc^{2}})^{-1}$ has been used for nuclear starburst such as SMGs and QSOs and the Milky Way value of $\alpha_{\rm CO}\sim4$ M$_{\odot}$ $({\rm K\,km\,s^{-1}\,pc^{2}})^{-1}$ in MS high redshift CSGs. Because of RCS0327 being cataloged as a starburst, to estimate the molecular mass we will use the values estimated for SMGs and starbursts at high redshift. 

\begin{figure*}[!t]
\epsscale{1.1}
\plotone{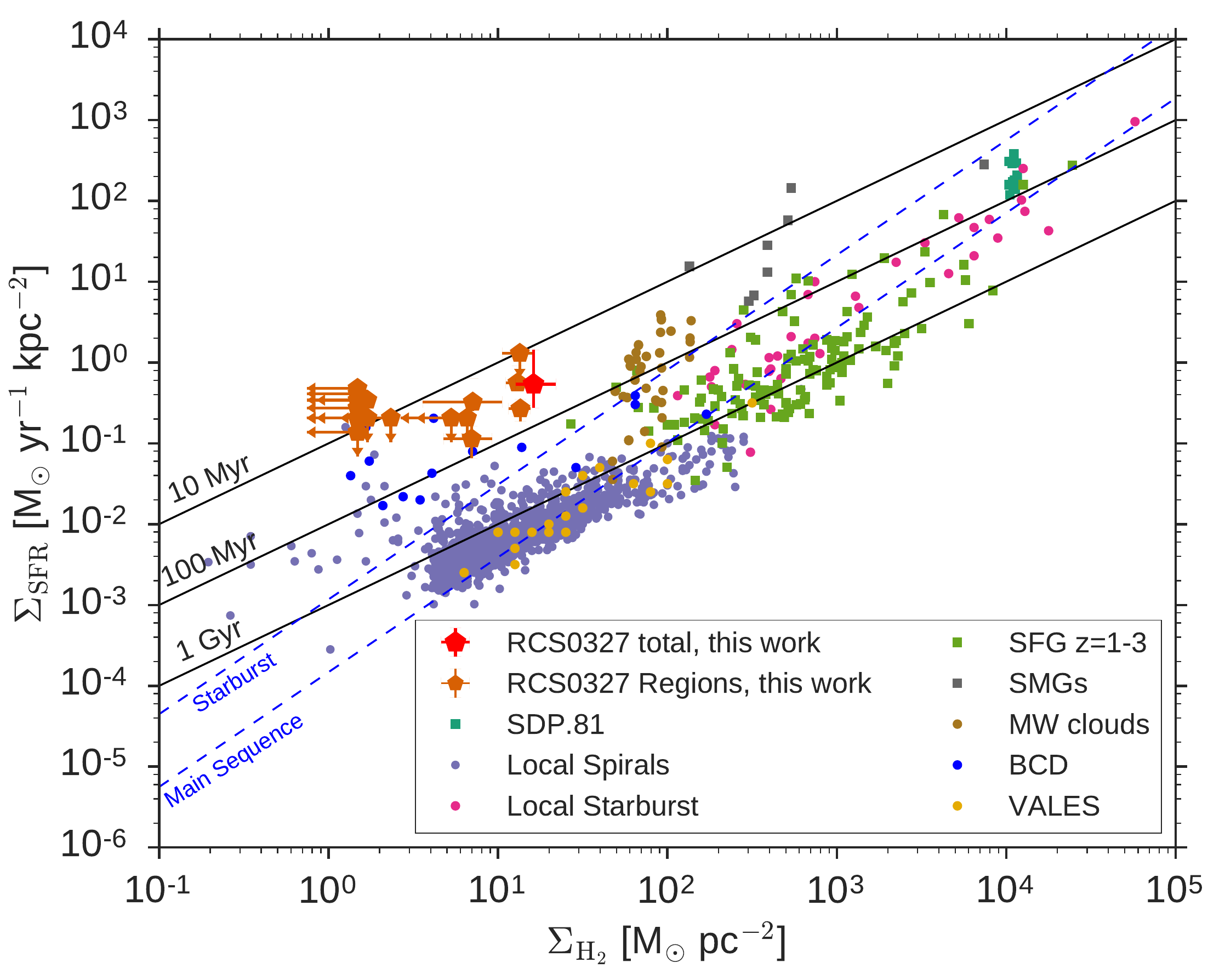}
\caption{Resolved properties (molecular gas and SFR surface density) for RCS0327 and a sample of other populations (References in the text). The red pentagon symbol shows the total properties found for RCS0327, while the orange pentagon symbols correspond to the different regions shown in Figure \ref{fig:source_plane}. The blue dashed lines show the relation found for MS and starburst galaxies \citep{Daddi2010}. The solid black lines show different depletion times assuming the available molecular gas fuel.  \label{fig:resolved_galaxies}}
\end{figure*}

Assuming a CO excitation level valid for SMGs of $L'_{\rm CO(3-2)}/L'_{\rm CO(1-0)}=0.66$ \citep[similar to $\sim0.6$ for MS CSGs,][]{Carilli_Walter2013} we obtain an intrinsic CO line luminosity of $L'_{\rm CO(1-0)}=4.40_{-0.35}^{+0.32}\times10^{8}$ ${\rm K\,km\,s^{-1}\,pc^{2}}$ and ${\rm H}_{2}=3.51_{-0.28}^{+0.26}\times10^{8}$ ${\rm M}_{\odot}$. Using the effective radius measured for the CO(3-2) emission we estimate the total area for the molecular gas surface of $A_{\rm CO(3-2)}=21.7\pm5.6$ kpc$^2$ and obtained a H2 surface density of $\Sigma_{\rm H2}=16.2_{-3.5}^{+5.8}\,{\rm M}_{\odot}\,{\rm pc}^{-2}$. 

In Figure \ref{fig:source_plane} we present the source plane emission from the optical, CO(3-2) and continuum at $450\,\mu m$. The regions where the CO(3-2) and continuum emission is produced corresponds to the clumps a--f \citep[see Figure 4 in][]{Sharon2012}, which have a combined spectral energy distribution of ${\rm SFR}=9.0_{-1.5}^{+8.3}\,{\rm M}_{\odot}\,{\rm yr}^{-1}$ \citep{Wuyts2014}. We use the estimated total ${\rm SFR}=29\pm8\,{\rm M}_{\odot}\,{\rm yr}^{-1}$ in the same region derived by \citet{Sharon2012} as an upper limit to the total SFR produced by clumps and the ISM combined. The SFR in combination with the molecular gas mass give a depletion time of $\sim40$ Myr and a SFR surface density of $\Sigma_{\rm SFR}=0.54_{-0.27}^{+0.89}\,{\rm M}_{\odot}\,{\rm yr}^{-1}\,{\rm kpc}^{-2}$.

We also derive resolved properties for individual regions plotted in the the right panel in Figure \ref{fig:source_plane}. These regions have a size of 0\farcs25 (2.2 kpc) in the source plane, which given the variable magnification some will be larger than the beam projected at the source plane. 
We use a set of 50 source plane reconstructions taken from the fitting iterations to estimate the significance of the detection in each region. We also take the image plane noise map into the source plane to give a proper upper limit for the emission in the regions with $S/N\leq2$. The results for all the regions are presented in Table \ref{tab:table}. 

In Figure \ref{fig:resolved_galaxies} we present the resolved properties for the total emission of RCS0327 and for each of the regions described above. We compare these results with those obtained for Milky Way molecular clouds \citep{Heiderman2010,Evans2014}, local spirals \citep{Bigiel2010,Kennicutt1998}, local starburst \citep{Kennicutt1998}, local blue compact dwarfs galaxies \citep[BCD,][]{Amorin2016}, low-redshift dusty normal star-forming galaxies \citep[VALES,][]{Villanueva2017}, $z=1-3$ star forming galaxies (SFG) \citep{Genzel2010,Tacconi2013,Freundlich2013}, SMGs \citep{Bothwell2010} and the lensed SMGs observed in high resolution by ALMA SDP.81 \citep{Hatsukade2015}. We notice that RCS0327 falls above the relation found by \citet{Daddi2010} for MS and starburst galaxies (blue dashed-lines), supporting the starburst nature of RCS0327. We point out that the position of RCS0327 on the diagram depends strongly on the assumed value for $\alpha_{\rm CO}$, but also even when using the Milky value of $\alpha_{\rm CO}\sim4$ M$_{\odot}$ $({\rm K\,km\,s^{-1}\,pc^{2}})^{-1}$ the galaxy would at most move to be on top of the starburst relation. 

For our estimate of the molecular mass we see that RCS0327 properties are similar to the local BCDs, which are low-metallicity starbursts galaxies showing higher star forming efficiencies when compared to normal disc galaxies \citep{Hunt2015,Amorin2016}. BCDs have already been found to work well as local analogs to similar redshift low-metallicity starbursts based on properties derived using optical spectroscopy \citep{Brammer2012}. Based on the latter, we can use the $\alpha_{\rm CO}$--metallicity relation found for BCDs to estimate an $\alpha_{\rm CO}$ value for RCS0327. The relation presented by \citet{Amorin2016} returns a value of $\alpha_{\rm CO}\sim25$ for RCS0327, consistent with the value given by other relations \citep{Hunt2015}.
The new $\alpha_{\rm CO}\sim25$ value would increase the estimate molecular gas mass for RCS0327 putting it near the MS relation with a depletion time of $\sim 1$ Gyr, in the same region as the $z=1-3$ SFGs. 

\subsection{CO excitation level}

We can use the detected CO(3-2) and CO(6-5) emission lines to constrain the CO excitation level in RCS0327. The total intrinsic flux densities return a fraction of $S_{\rm CO(6-5)}/S_{\rm CO(3-2)}=1.1\pm0.2$. This value is consistent with having the peak at ${\rm J}\sim5$, similar to some CO excitation levels measured on SMGs and lower than the excitation level measured for QSOs \citep{Carilli_Walter2013}.

We can use the same method presented in the previous section to obtain resolved measurements of $S_{\rm CO(6-5)}/S_{\rm CO(3-2)}$ in the regions presented in Figure \ref{fig:source_plane}. We have 5 regions (see Table \ref{tab:table}) with detections of CO(3-2) and constraints on CO(6-5). The ratios go from $S_{\rm CO(6-5)}/S_{\rm CO(3-2)}<0.3$ in region 10 to $S_{\rm CO(6-5)}/S_{\rm CO(3-2)}=3.0\pm0.9$ in region 9. Our results are consistent with those found for the high-z disk galaxy presented by \citet{Bournaud2015}, where the CO excitation level is $S_{\rm CO(6-5)}/S_{\rm CO(3-2)}>1$ for the main star-forming clumps and $S_{\rm CO(6-5)}/S_{\rm CO(3-2)}<1$ for the inter-clump gas. The warmer and denser gas associated with the main clumps allows for higher CO excitation level when compared to the more extended gas. In the case of RCS0327, the main star-forming clumps identified by \citet{Sharon2012} and \citet{Wuyts2014} correspond to the regions 2 and 9, which have $S_{\rm CO(6-5)}/S_{\rm CO(3-2)}$ ratios of $1.3\pm0.6$ and $3.0\pm0.9$. The regions 1, 3 and 10 show $S_{\rm CO(6-5)}/S_{\rm CO(3-2)}<1$ values, consistent with the inter-clump gas of the simulations (Figure \ref{fig:source_plane}). 

\section{Conclusion}

We have presented ALMA observations of the emission lines CO(3-2), CO(6-5) and $450\,\mu m$ rest-frame continuum emission in the $z=1.7$ young low-metallicity starburst strongly-lensed galaxy RCS0327. The source plane reconstruction of the detected emission reveals that the molecular gas, traced by CO(3-2), is located on top of the star-forming clumps showing the gas reservoir fueling the ongoing star-forming process. The continuum dust emission follows a similar angular extension and distribution as the molecular gas but also extending towards parts of the galaxy not as bright in CO(3-2), showing a clear spatial offset. 
The molecular gas and SFR surface density show that RCS0327 is a low-density starburst, similar to local BCDs and probably triggered by an ongoing merger. These results support the scenario where BCDs are identified as local counterpart to the low-metallicity starburst galaxies at high redshift. 

The detected CO(3-2) and CO(6-5) return a CO excitation level consistent with having the  peak at ${\rm J}\sim5$ at large scales. The total excitation appears to be the result of the combination of higher excitation regions near the star-forming clumps and lower excitation regions over the more extended gas phase. This is one of the first times where the CO excitation level is resolved in detail on a galaxy at high redshift. 

We have shown that giant gravitational arcs offer an excellent opportunity to resolve in detail the different phases of the ISM in the cases where a good lensing model is in hand. Our coarse observations already show that RCS0327 is not well described by a single mode star-forming galaxy, showing different CO excitation levels, molecular gas reservoirs and dust obscuration across $\sim8\,{\rm kpc}$. Future high angular resolution observations and the extension to other bright arcs will take us one step closer to understanding the star-formation process in galaxies at high redshift.

\acknowledgments
This paper makes use of the following ALMA data: ADS/JAO.ALMA\#2015.1.00920.S.
ALMA is a partnership of ESO (representing its member states), NSF (USA) and
NINS (Japan), together with NRC (Canada) and NSC and ASIAA (Taiwan), in
cooperation with the Republic of Chile. The Joint ALMA Observatory is operated
by ESO, AUI/NRAO and NAOJ. 
This research has been supported by CONICYT-Chile grant Basal-CATA PFB-06/2007, FONDECYT Regular 1141218 and ALMA-CONICYT project 31160033.
M.A. acknowledges partial support from FONDECYT through grant 114009.

\end{document}